\documentclass[12pt]{iopart}

\usepackage{graphicx}

\begin{document}

\title[Growth signals determine the topology of evolving networks]{Growth signals determine the topology of evolving networks}

\author{Ana Vrani\'c}
\address{Institute  of  Physics  Belgrade, University of Belgrade, Belgrade}
\ead{anav@ipb.ac.rs}

\author{Marija Mitrovi\'c Dankulov}

\address{Institute  of  Physics  Belgrade, University of Belgrade, Belgrade}
\ead{mitrovic@ipb.ac.rs}
\vspace{10pt}
\begin{indented}
\item[]September 2020
\end{indented}

\begin{abstract}
Network science provides an indispensable theoretical framework for
studying the structure and function of real complex systems. Different
network models are often used for finding the rules that govern their
evolution, whereby the correct choice of model details is crucial for
obtaining relevant insights. We here study how the structure of
networks generated with the aging nodes model depends on the
properties of the growth signal. We use different fluctuating signals
and compare structural dissimilarities of the networks with those
obtained with a constant growth signal. We show that networks with
power-law degree distributions, which are obtained with time-varying
growth signals, are correlated and clustered, while networks obtained
with a constant growth signal are not. Indeed, the properties of the
growth signal significantly determine the topology of the obtained
networks and thus ought to be considered prominently in models of
complex systems.

\end{abstract}

%
%
%
%
%

\section{Introduction}

Emergent collective behavior is an indispensable property of complex systems \cite{ladyman2013}. It occurs as a consequence of interactions between a large number of units that compose a complex system, and it cannot be easily predicted from the knowledge about the behavior of these units. The previous research offers a definite proof that the structure of the interaction network is inextricably associated with dynamic and function of the complex system \cite{barrat2008, pascual2006, castellano2009, gosak2018, arenas2008, boccaletti2016, chen2018, kuga2018}. The structure of complex networks is essential for understanding the evolution and function of various complex systems \cite{boccaletti2006,newman2010, holme2012, boccaletti2014}.  

The structure and dynamics of real complex systems are studied using complex network theory \cite{boccaletti2006,newman2010, ladyman2013}. It was shown that real networks have similar topological properties regardless of their origins \cite{barabasi2009}. They have broad degree distribution, degree-degree correlations, and power-law scaling of clustering coefficient \cite{barabasi2009,newman2010}. Understanding how these properties emerge in complex networks leads to the factors that drive their evolution and shape their structure \cite{barrat2008}.

The complex network models substantially contribute to our understanding of the connection between the network topology and system dynamics and uncover underlying mechanisms that lead to the emergence of distinctive properties in real complex networks \cite{barabasi1999, tadic2001, mitrovic2009}. For instance, the famous Barabasi-Albert model \cite{barabasi1999} finds the emergence of broad degree distribution to be a consequence of preferential attachment and network growth. Degree-degree anti-correlations of the Internet can be explained, at least to a certain extent, by single edge constrain \cite{maslov2004, park2003}. Detailed analysis of emergence of clustered networks shows that clustering is either the result of finite memory of the nodes \cite{klemm2002} or occurs due to triadic closure \cite{serrano2005}. 

Network growth, in combination with linking rules, shapes the network topology \cite{vazquez2003}. While various rules have been proposed to explain the topology of real networks \cite{boccaletti2006}, most models assume a constant rate of network growth, i.e., the addition of a fixed number of nodes at each time step \cite{barabasi1999, klemm2002, serrano2005}. However, the results of empirical analysis of numerous technological and social systems show that their growth is time-dependent \cite{huberman1999, mitrovic2010a, mitrovic2015, liu2019}. The accelerated growth in complex networks is the cause of the high heterogeneity in the distribution of webpages among websites \cite{huberman1999} and the emergence of highly cited authors in citation networks \cite{liu2019}. The growth of real systems is not always accelerated. The number of new nodes joining the system varies in time, has trends, and exhibits circadian cycles typical for human behavior \cite{mitrovic2010a, mitrovic2012, mitrovic2015}. These signals are multifractal and have long-range correlations \cite{kantelhardt2002}. Some preliminary evidence shows that the time-varying growth influences the structure and dynamics of the social system and, consequently, the structure of interaction networks in social systems \cite{mitrovic2012,mitrovic2011,mitrovic2015,tadic2017,tadic2013}. Still, which properties of the real growth signal have the largest influence, how different properties influence the topology of the generated networks and to what extent is an open question.

In this work, we explore the influence of real and computer-generated time-varying growth signals on complex networks' structural properties. We adapt the aging nodes model \cite{hajra2004} to enable time-varying growth. We compare the structure of networks generated using the growing signals from empirical data and randomized signals with ones grown with the constant signal using D-measure \cite{tiago2}. We demonstrate that the growth signal determines the structure of generated networks. The networks grown with time-varying signals have significantly different topology compared to networks generated through constant growth. The most significant difference between topological properties is observed for the values of model parameters for which we obtain networks with broad degree distribution, a common characteristic of real networks \cite{boccaletti2006}. Our results show that real signals, with trends, cycles, and long-range correlations, alter the structure of networks more than signals with short-range correlations.

This paper is divided as follows. In section \ref{sec:2}, we provide a detailed description of growth signals. In section \ref{sec:3}, we briefly describe the original model with aging nodes and structural properties of networks obtained for different values of model parameters \cite{hajra2004}. We also describe the changes in the model that we introduce to enable time-varying growth. We describe our results in the section \ref{sec:4} and show that the values of D-measure indicate large structural differences between networks grown with fluctuating and ones grown with constant signals. This difference is particularly evident for networks with power-law degree distribution and real growth signals. The networks generated with real signals are correlated and have hierarchical clustering, properties of real networks that do not emerge if we use a constant growth. We discuss our results and give a conclusion in section \ref{sec:5}.

\section{Growth signals \label{sec:2}}

The \textit{growth signal} is the number of new nodes added in each time step. Real complex networks evolve at a different pace, and the dynamics of link creation define the time unit of network evolution. For instance, the co-authorship network evolves through establishing a link between two scientists when they publish a paper \cite{sarigol2014}, while the links in an online social network are created at a steady pace, often interrupted by sudden bursts \cite{myers2014}. A publication of a paper is thus a unit of time for the evolution of co-authorship networks, while the most appropriate time unit for social networks is one minute or one hour. While networks may evolve at a different pace, their evolution is often driven by the related mechanisms reflected by the similarity of their structure \cite{boccaletti2006}.

\begin{figure}[!ht]
\centering
\includegraphics[width=1\textwidth]{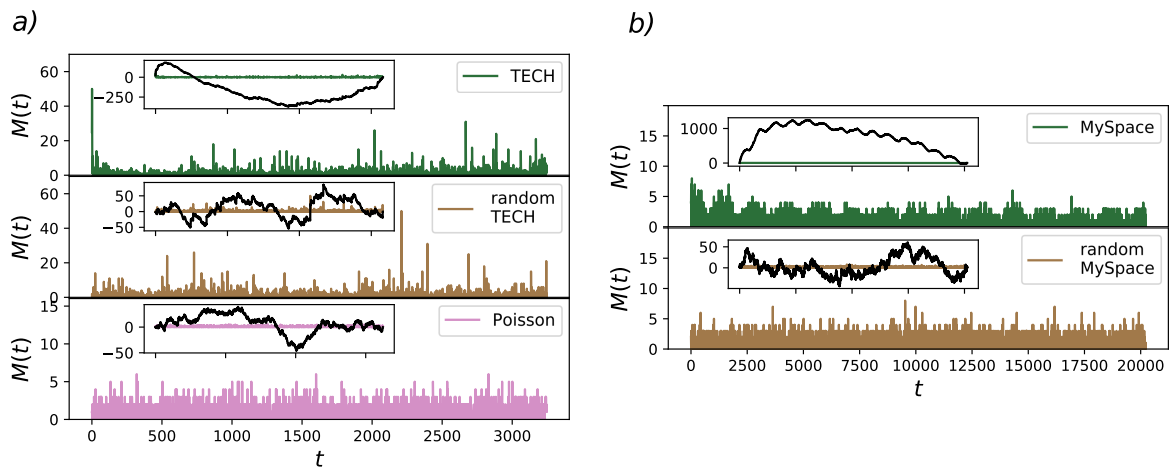}
\caption{Growth signals for TECH (a) and MySpace (b) social groups, their randomized counterparts, and random signal drawn from Poasonian distribution with mean $1$. The cumulative signals are shown in insets.}
\label{fig:signals}
\end{figure}

In this work, we use two different growth signals from real systems figure \ref{fig:signals}: (a) the data set from TECH community from Meetup social website \cite{smiljanic2017}  and (b) two months dataset of MySpace social network \cite{suvakov2013}. TECH is an event-based community where members organize offline events through the Meetup site \cite{smiljanic2017}. The time unit for TECH is event since links are created only during offline group meetings. The growth signal is the number of people that attend the group's meetings for the first time. MySpace signal shows the number of new members occurring for the first time in the dataset \cite{suvakov2013} with a time resolution of one minute. The number of newly added nodes for the TECH signal is $N=3217$, and the length of the signal is $T=3162$ steps. We have shortened the MySpace signal to $T=20221$ time steps to obtain the network with $N=10 000$ nodes.

Real growth signals have long-range correlations, trends and cycles \cite{suvakov2013,mitrovic2012,mitrovic2015}. We also generate networks using randomized signals and one computer-generated white-noise signal to explore the influence of these signal's features on the structure of evolving networks. We randomize real signals using reshuffling procedure and keep their length and mean value, the number of added nodes, and probability density function of fluctuations intact, but destroy cycles, trends, and long-range correlations. Besides, we generate a white-noise signal from a Poissonian probability distribution with a mean equal to $1$. The length of the signal is $T=3246$, and the number of added nodes in the final network is the same as for the TECH signal. 

We characterize the long-range correlations of the growth signals calculating Hurst exponent \cite{peng1994, kantelhardt2001}. Hurst exponent describes the scaling behavior of time series $M(xt)=x^{H}M(t)$. It takes values between $0.5$ and $1$ for long-range correlated signals and $H=0.5$ for short-range correlated signals. The most commonly used method for estimating Hurst exponent of real, often non-stationary, temporal signals is detrended fluctuation analysis (DFA) \cite{peng1994}. The DFA removes trends and cycles of real signals and estimates Hurst exponent based on residual fluctuations. The DFA quantifies the scaling behaviour of the second moment fluctuations. However, signals can have deviations in fractal structure with large and small fluctuations that are characterized by different values of Hurst exponents \cite{kantelhardt2002}. 

We use multifractal detrended fluctuation analysis (MFDFA) \cite{kantelhardt2002, ihlen2012} to estimate multifractal Hurst exponent H(q). For given time series $\{x_i\}$ with length N, first we define global profile in the form of cumulative sum equation \ref{eq:cumsum}, where where $\langle x\rangle $ represents average of the time series:
\begin{equation}
Y(j) = \sum_{i=0} ^j (x_i - \langle x\rangle), \quad j=1, ..., N
\label{eq:cumsum}
\end{equation}

Subtracting the mean of the time series is supposed to eliminate global trends. Insets of figure \ref{fig:signals} show global profiles of TECH, MySpace, their randomized signals and Poasonian distribution. The profile of the signal Y is divided into $N_s = int (N/s)$ non overlapping segments of length s. If $N$ is not divisible with s the last segment will be shorter. This is handled by doing the same division from the opposite side of time series which gives us $2N_s$ segments. From each segment $\nu$, local trend $p^m_{\nu, s}$ - polynomial of order m - should be eliminated, and the variance $F^2(\nu, s)$ of detrended signal is calculated as in equation \ref{eq:var}:
\begin{equation}
F^2(\nu, s) = \frac{1}{s}\sum_{j=1}^s \left[Y(j) - p^m_{\nu, s}(j)\right]^2
\label{eq:var}
\end{equation}
Then the q-th order fluctuating function is: 
\begin{equation}
\eqalign{F_q(s) = &\left\{\frac{1}{2N_s}\sum_{\nu}^{2N_s}\left[F^2(\nu, s)\right]^{\frac{q}{2}}\right\}^{\frac{1}{q}},  q \neq 0 \nonumber \\
        F_0(s) = &\exp \left\{\frac{1}{4N_s}\sum_{\nu}^{2N_s}ln \left[F^2(\nu, s)\right]\right\}, q=0} 
\end{equation}

The fluctuating function scales as power-law $F_q(s) \sim s^{H(q)}$ and the analysis of log-log plots $F_q(s)$ gives us an estimate of multifractal Hurst exponent $H(q)$. Multifractal signal has different scaling properties over scales while monofractal is independent of the scale, i.e., H(q) is constant. 

\begin{figure}[!ht]
\centering
\includegraphics[width=0.6\linewidth]{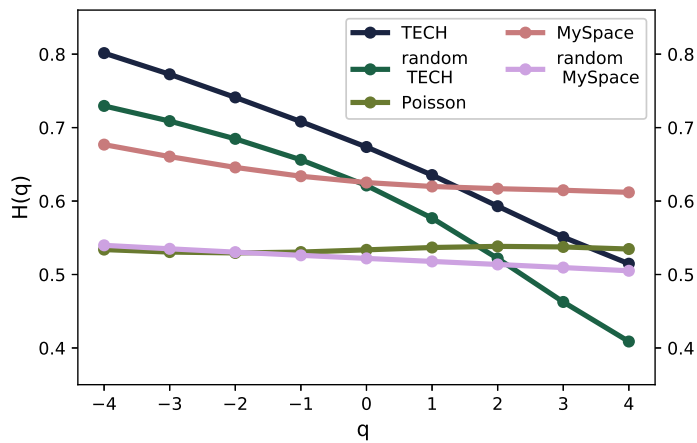}
\caption{Dependence of Hurst exponent on parameter $q$ for all five signals shown in figure \ref{fig:signals} obtained with MFDFA. }
\label{fig:mfdfa}
\end{figure}

Figures \ref{fig:signals} (a) and \ref{fig:mfdfa} show that the TECH signal has long trends and a broad probability density function of fluctuations. The trends are erased from the randomized TECH signal, but the broad distribution of the signal and average value remain intact. MFDFA analysis shows that real signals have long-range correlations with Hurst exponent approximately $0.6$ for $q=2$, figure \ref{fig:mfdfa}. The TECH signal is multifractal, the consequence of both broad probability distribution for the values of time series and different long-range correlations of the intervals with small and large fluctuations. Shuffling of the time series does not destroy the broad distribution of values, the reason for the persistent multifractality of the TECH randomized signal, figure \ref{fig:mfdfa}.

MySpace signal has a long trend with additional cycles that are a consequence of human circadian rhythm, figure \ref{fig:signals}(b). It is multifractal for $q<0$, and has constant value of $H(q)$ for $q>0$, figure \ref{fig:mfdfa}. In MFDFA, with negative values of $q$, we put more emphasis on segments with smaller fluctuations, while for positive $q$ emphasis is more on segments with larger fluctuations \cite{ihlen2012}. Segments with smaller fluctuations have more persistent long-range correlations in both real signals, see figure \ref{fig:mfdfa}. Randomized MySpace signal and Poissonian signal are monofractal and have short-range with $H=0.5$ correlations typically for white noise.    

\section{Model of ageing nodes with time-varying growth \label{sec:3}}

To study the influence of temporal fluctuations of growth signal on network topology, we need a model with linking rules where linking probability between network nodes depends on time. We use a network model with aging nodes \cite{hajra2004}. In this model, the probability to link the newly added node and the old one is proportional to their age difference and the degree of an old node. In the original version of the model, one node is added to the network and linked to one old node in each time step. The old node is chosen according to probability
\begin{equation}
\Pi_{i}(t)\sim k_{i}(t)^{\beta}\tau_{i}^{\alpha} 
\label{eq:1}
\end{equation}
where $k_{i}(t)$ is a degree of a node $i$ at time $t$, and $\tau_{i}$ is age difference between node $i$ and newly added node. As it was shown in reference \cite{hajra2004}, the values of model parameters $\beta$ and $\alpha$ determine the topological properties of the resulting networks grown with the constant signal. According to this work, the networks generated using constant growth signals are uncorrelated trees for all values of model parameters. The phase diagram in $\alpha-\beta$ plain, obtained for $\beta>0$ and $\alpha<0$, shows that the degree distribution $P(k) \sim k^{-\gamma}$ with $\gamma=3$ is obtained only along the line $\beta(\alpha^{*})$, see reference \cite{hajra2004}. For $\alpha>\alpha^{*}$ networks have gel-like small world behavior, while for $\alpha<\alpha^{*}$ but close to line $\beta(\alpha^{*})$ networks have stretched exponential shape of degree distribution \cite{hajra2004}. 

Here we slightly change the original aging model \cite{hajra2004} to enable the addition of more than one node and more than one link per newly added node in each time step. In each time step, we add $M\geq1$ new nodes to the network and link them to $L\geq1$ old nodes according to probability $\Pi_i$ given in equation \ref{eq:1}. Again, the networks with broad degree distribution are only generated for the combination of the model parameters along the critical line $\beta(\alpha^{*})$. The position of this line in the $\alpha-\beta$ plane changes with link density, while the addition of more than one node in each time step does not influence its position. Our analysis shows that the position of the critical line is independent of the growth signal's properties, see figure \ref{fig:dmeasure}. Networks obtained for the values of model parameters $\beta(\alpha^{*})$, $L\geq 2$, and constant growth have power-law degree distribution, are uncorrelated and have a finite non-zero value of clustering coefficient which does not depend on node degree (see figure \ref{fig:properties}(b)). For $\alpha<\alpha^{*}$ we obtain networks with stretched exponential degree distribution, without degree-degree correlations and small value of clustering exponent that does not depend on node degree (see figure \ref{fig:properties}(a)). Correlated networks with power-law dependence of the clustering coefficient on the degree are generated for $L\geq2$, $\alpha>\alpha^{*}$, and constant growth, see figure \ref{fig:properties}. However, these networks do not have a power-law degree distribution. 

\section{Structural differences between networks generated with different growth signals \label{sec:4}}

We generate networks for different values of $L$ and different growth signal profiles $M(t)$. To examine how these properties influence the network structure, we compare the structure of networks obtained with different growth signals with networks of the same size grown with constant signal $M=1$. The $M=1$ is the closest constant value to average values of the signals, which are $1.017$ for TECH, $0.47$ for MySpace, and $1$ for Poissonian signals. We explore the parameter space of the model by generating networks for pairs of values $(\alpha,\beta)$ in the range $-3\leq\alpha\leq-0.5$ and $1\leq\beta\leq3$ with steps $0.5$. For each pair of $(\alpha,\beta)$ we generated networks of different link density by varying parameter $L\in{1,2,3}$, and for each combination of $(\alpha, \beta, L)$, we generate a sample of $100$ networks and compare the structure of the networks grown with $M=1$ with the ones grown with $M(t)$ shown in figure \ref{fig:signals}. 

\begin{figure}[ht]
    \centering
    \includegraphics[width=0.6\linewidth]{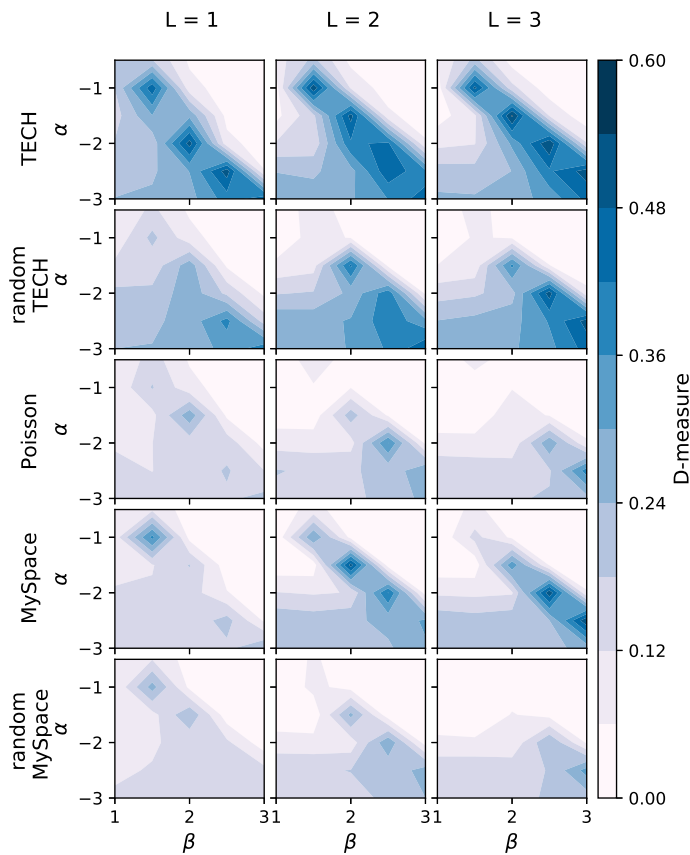}
    \caption{The comparison of networks grown with growth signals shown in figure \ref{fig:signals} versus ones grown with constant signal $M=1$, for value of parameter $\alpha\in[-3,-1]$ and $\beta\in[1,3]$. $M(t)$ is the number of new nodes, and $L$ is the number of links added to the network in each time step. The compared networks are of the same size.}
    \label{fig:dmeasure}
\end{figure}

We quantify topological differences between two networks using D-measure defined in reference \cite{tiago2}
\begin{equation} \label{eq:2}
\eqalign{
D(G,G') =\omega \left| \sqrt{\frac{J(P_1,..P_N)}{log(d)}}-\sqrt{\frac{J(P_1^{'},..P_N^{'})}{log(d^{'})}} \right| \nonumber +  (1-\omega) \sqrt{\frac{J(\mu_{G},\mu_{G^{'}})}{log2}}.}
\end{equation}

D-measure captures the topological differences between two networks, $G$ and $G',$ on a local and global level. The first term in equation \ref{eq:2} evaluates dissimilarity in average node connectivity as a difference between Jensen-Shannon divergences $J(P_{1},\ldots, P_{N})$ for node distance distributions $P_{1}, \ldots, P_{N}$. The second term of D-measure compares averaged nodes distance distributions, $\mu_{G}$ and $\mu_{G^{'}}$ through Jensen-Shannon divergence, capturing global differences between two networks. The original definition of D-measure also includes the third term, which quantifies dissimilarity in node $\alpha$-centrality. The term can be omitted without precision loss \cite{tiago2}. By definition, the importance of differences between local and global properties of two networks can be tuned in D-measure by setting the value of parameter $\omega$. Since we are interested in the overall topological difference between two networks, we consider global and local features to be equally important and set $\omega=0.5$. The D-measure takes the value between $0$ and $1$. The lower value of D-measure is the more similar two networks are, with $D=0$ for isomorphic graphs. The D-measure outperforms previously used measures of network dissimilarity such as Hamming distance and graph editing distance and clearly distinguish between networks generated with the same model but with different values of model parameters \cite{tiago2}.

For each pair of networks, one grown with constant and one with the fluctuating signal, we calculate the D-measure. The structural difference between networks grown with constant and fluctuating growth signal for fixed $L$ and values of parameters $\alpha$ and $\beta$ is obtained by averaging the D-measure calculated between all possible pairs of networks, see figure \ref{fig:dmeasure}. We observe the non-zero value of D-measure for all time-varying signals. The D-measure has the largest value in the region around the line $\beta(\alpha^{*})$. The values of D-measure in this region are similar to ones observed when comparing Erd\"{o}s–R\'{e}nyi graphs grown with linking probability below and above critical value \cite{tiago2}. For values $\beta<\beta(\alpha^{*})$, the structural differences between networks grown with constant signal and $M(t)$ still exist, but they become smaller as we are moving away from the critical line. Networks obtained with constant signal and fluctuating signals have statistically similar structural properties in the region of small-world network gels, i.e., $\alpha>\alpha^{*}$.

\begin{figure}[h!]
    \centering
    \includegraphics[width=0.85\textwidth]{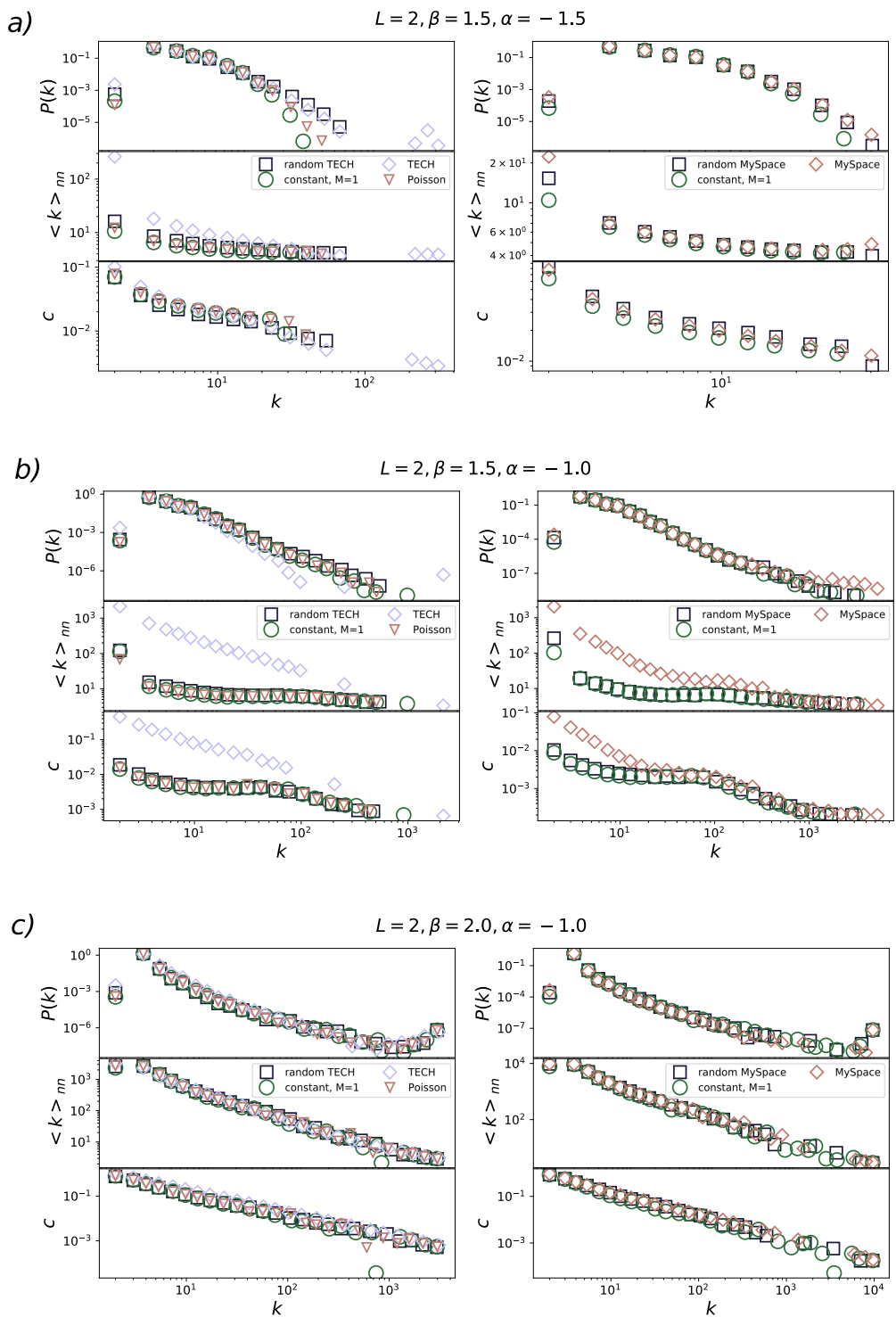}

    \caption{Degree distribution, the dependence of average first neighbor degree on node degree, dependence of node clustering on node degree for networks grown with different time-varying and constant signals. Model parameters have value $\alpha=-1.5$, $\beta=1.5$ (a), $\alpha=-1.0$, $\beta=1.5$ (b), $\alpha=-1.0$, $\beta=2.0$ (c), and $L=2$ for all networks.}
    \label{fig:properties}
\end{figure}

We focus on the region around the critical line and observe the significant structural discrepancies between networks created for constant versus time-dependent growth signals for all signals regardless of their features. However, the value of D-measure depends on the signal's properties, figure \ref{fig:dmeasure}. Networks grown with multifractal signals, TECH, random TECH, and MySpace signals, are most different from those created by a constant signal. The D-measure has the maximum value for the original TECH signal, with $D_{max}=0.552$, the signal with the most pronounced multifractal properties among all signals shown in figure \ref{fig:mfdfa}. Networks generated with randomized MySpace signal and Poisson signal are the least, but still notably dissimilar from those created with $M=1$. 

The value of D-measure rises with a decline of $\alpha^{*}$. This observation can be explained by a closer examination of linking rules and how model parameters determine linking dynamics between nodes. The ability of a node to acquire a link declines with its age and grows with its degree. A node's potential to become a hub, nodes with the degrees that are significantly larger than average network degree, depends on the number of nodes added to the network in the $T$ time steps after its birth. The length of the interval $T$ decreases with parameter $\alpha$. In the case of constant signal, the number of nodes added during this time interval is constant and equal to $MT$. For fluctuating growth signals, the number of added nodes during the time $T$ varies with time. In signals that have a broad distribution of fluctuations, like TECH signals, the peaks of the number of newly added nodes lead to the emergence of one or several hubs and super hubs. The emergence of super hubs, nodes connected to more than $30\%$ of the nodes in the network, significantly alters the network's topology. For instance, the existence of super hubs lowers the value of average path length and network diameter \cite{boccaletti2006}. The emergence of hubs occurs for values of parameter $\alpha$ relative close to $-1.0$ for signals with long-range correlations. As we decrease the parameter $\alpha$, the fluctuations present in the time-varying signals become more important, and we observe the emergence of hubs even for the white-noise signals. The trends present in real growth signals further promote the emergence of hubs. The impact of fluctuations and their temporal features on the structure of complex networks increases with link density. 

The large number of structural properties observed in real networks are often consequences of particular degree distributions, degree correlations, and clustering coefficient \cite{orsini2015}. Figure \ref{fig:properties} shows degree distribution $P(k)$, dependence of average neighbouring degree on node degree $\langle k\rangle_{nn}(k)$, and dependence of clustering coefficient on node degree $c(k)$ for networks with average number of links per node $L=2$. The significant structural differences between networks grown with real time-varying and constant signals are observed for the values of model parameters $\alpha=-1.0$ and $\beta=1.5$, figure \ref{fig:dmeasure} and figure \ref{fig:properties}(b). The degree distribution of networks generated for real signals shows the occurrence of super hubs in these networks, while degree distributions of networks generated with white-noise like signals do not differ from one created with constant signal figure \ref{fig:properties}(b). Networks obtained for the real signals are disassortative and have a hierarchical structure, i.e., their clustering coefficient decreases with the degree. On the other hand, networks generated with constant and randomized signals are uncorrelated, and their clustering weakly depends on the degree. 

We observe a much smaller, but still noticeable, difference between the topological properties of networks evolved with constant and time-varying signal for $\alpha<\alpha^{*}$, figure \ref{fig:properties}(a). The difference is particularly observable for degree distribution and dependence of average neighboring degree on node degree of networks grown with real TECH signal. The fluctuations of time-varying growth signals do not influence the topological properties of small-world gel networks, figure \ref{fig:properties}(c). For $\alpha>\alpha^{*}$, the super hubs emerge even with the constant growth. Since this is the mechanism through which the fluctuations alter the structure of evolving networks for $\alpha\leq\alpha^{*}$, the features of the growth signals cease to be relevant.  

\section{ Discussion and Conclusions \label{sec:5}}

We demonstrate that the resulting networks' structure depends on the features of the time-varying signal that drives their growth. The previous research \cite{mitrovic2012,mitrovic2015} indicated the possible influence of temporal fluctuations on network properties. Our results show that the temporal properties of growth signals generate networks with power-law degree distribution, non-trivial degree-degree correlations, and clustering coefficient even though the local linking rules, combined with constant growth, produce uncorrelated networks for the same values of model parameters \cite{hajra2004}. 

We observe the most substantial dissimilarity in network structure along the critical line, the values of model parameters for which we generate networks with broad degree distribution. Figure \ref{fig:dmeasure} shows that dissimilarity between networks grown with time-varying signals and ones grown with constant signals always exists along this line regardless of the features of growth signal. However, the magnitude of this dissimilarity strongly depends on these features. We observe the largest structural difference between networks grown with multifractal TECH signal and networks that evolve by adding one node in each time step. The identified value of D-measure is similar to one calculated in the comparison between sub-critical and super-critical Erd\"{o}s–R\'{e}nyi graphs \cite{tiago2} indicating the considerable structural difference between these networks. Our findings are further confirmed in figure \ref{fig:properties}(b). The networks generated with signals that have trends and long-range temporal correlations differ the most from those grown with the constant signal. Our results show that even white-noise type signals can generate networks significantly different from ones created with constant signal for low values of $\alpha^{*}$.

The value of D-measure declines fast as we move away from the critical line, figure \ref{fig:dmeasure}. The main mechanism through which the fluctuations influence the structure of evolved networks is the emergence of hubs and super hubs. For values of $\alpha<<\alpha^{*}$, the nodes attache to their immediate predecessors creating regular networks without hubs. For $\alpha \sim \alpha^{*}$ graphs have stretched exponential degree distribution with low potential for the emergence of hubs. Still, multifractal signal TECH enables the emergence of hub even for the values of parameters for which we observe networks with stretched-exponential degree distribution in the case of constant growth figure \ref{fig:properties}(a). By definition, small-world gels generated for $\alpha>\alpha^{*}$ have super-hubs \cite{hajra2004} regardless of the growth signal, and therefore the effects that fluctuations produce in the growth of networks do not come to the fore for values of model parameters in this region of $\alpha-\beta$ plane.

Evolving network models are an essential tool for understanding the evolution of social, biological, and technological networks and mechanisms that drive it \cite{boccaletti2006}. The most common assumption is that these networks evolve by adding a fixed number of nodes in each time step \cite{boccaletti2006}. So far, the focus on developing growing network models was on linking rules and how different rules lead to networks of various structural properties \cite{boccaletti2006}. Growth signals of real systems are not constant \cite{mitrovic2015,mitrovic2012}. They are multifractal, characterised with long-range correlations \cite{mitrovic2015}, trends and cycles \cite{suvakov2013}. Research on temporal networks has shown that temporal properties of edge activation in networks and their properties can affect the dynamics of the complex system \cite{holme2012}. Our results imply that modeling of social and technological networks should also include non-constant growth and that its combination with local linking rules can significantly alter the structure of generated networks.

\ack
We acknowledge funding provided by the Institute of Physics Belgrade, through the grant by the Ministry of Education, Science, and Technological Development of the Republic of Serbia.  Numerical simulations were run on the PARADOX-IV supercomputing facility at the Scientific Computing Laboratory, National Center of Excellence for the Study of Complex Systems, Institute of Physics Belgrade. The work of M.M.D. was, in part, supported by the Ito Foundation fellowship.

\end{document}